\documentclass[times,twocolumn,final]{elsarticle}

\usepackage{framed,multirow}

\usepackage{amssymb}
\usepackage{latexsym}

\usepackage{url}
\usepackage{amsmath}
\usepackage{hyperref}
\usepackage{dsfont}
\usepackage{upgreek}
\usepackage{rotating}
\usepackage{graphicx} 

\makeatletter

\begin{document}


\begin{frontmatter}

\title{MONAI Label: A framework for AI-assisted Interactive Labeling of 3D Medical Images}%


\author[kcl,nvidia]{Andres~Diaz-Pinto\corref{cor1}} \ead{andres.diaz-pinto@kcl.ac.uk}
\author[nvidia]{Sachidanand~Alle}
\author[nvidia]{Vishwesh~Nath}
\author[nvidia]{Yucheng~Tang}
\author[nvidia]{Alvin~Ihsani}
\author[kcl]{Muhammad~Asad}
\author[kcl,ucl]{Fernando~P\'{e}rez-Garc\'{i}a}
\author[kcl,ucl]{Pritesh~Mehta}
\author[nvidia]{Wenqi~Li}
\author[nvidia]{Mona~Flores}
\author[nvidia]{Holger R.~Roth}
\author[kcl]{Tom~Vercauteren}
\author[nvidia]{Daguang Xu}
\author[nvidia]{Prerna Dogra}
\author[kcl]{Sebastien~Ourselin}
\author[nvidia]{Andrew~Feng}
\author[kcl]{M. Jorge~Cardoso}

\cortext[cor1]{Corresponding author}
\address[kcl]{School of Biomedical Engineering \& Imaging Sciences, King’s College London. London, UK.}
\address[nvidia]{NVIDIA Santa Clara, CA, USA.}
\address[ucl]{Department of Medical Physics and Biomedical Engineering, University College London. London, UK.}



\begin{abstract}
The lack of annotated datasets is a major bottleneck for training new task-specific supervised machine learning models, considering that manual annotation is extremely expensive and time-consuming. To address this problem, we present MONAI Label, a free and open-source framework that facilitates the development of applications based on artificial intelligence (AI) models that aim at reducing the time required to annotate radiology datasets. Through MONAI Label, researchers can develop AI annotation applications focusing on their domain of expertise. It allows researchers to readily deploy their apps as services, which can be made available to clinicians via their preferred user interface. Currently, MONAI Label readily supports locally installed (3D Slicer) and web-based (OHIF) frontends and offers two active learning strategies to facilitate and speed up the training of segmentation algorithms. MONAI Label allows researchers to make incremental improvements to their AI-based annotation application by making them available to other researchers and clinicians alike. Additionally, MONAI Label provides sample AI-based interactive and non-interactive labeling applications, that can be used directly off the shelf, as plug-and-play to any given dataset. Significant reduced annotation times using the interactive model can be observed on two public datasets.
\end{abstract}


\end{frontmatter}


\section*{Introduction}
\label{introduction}

Image segmentation and quantitative analysis of medical image data, especially of 3D volumes, are tedious and time-consuming tasks. The lack of expert-annotated datasets is one of the primary bottlenecks when developing new supervised segmentation deep learning algorithms. Non-expert manual annotation is time-consuming, cost-ineffective, and often even with trained groups of annotators, resulting labels have variance. Conversely, expert-led annotations (e.g. clinicians, anatomists) result in higher quality labels, but due to time, availability and cost, the number and variety of annotated samples suffer.


Deep learning algorithms are the current state of the art for automatic 2D and 3D medical image segmentation \cite{Isensee2020, He2021, Hatamizadeh2021}, inspired by the landmark contributions of \cite{Ronneberger2015} (2D U-Net), \cite{Cicek2016} (3D U-Net), and \cite{Milletari2016} (V-Net). The aforementioned models are the standard benchmark models on the medical segmentation decathlon challenge (\textit{MSD}) \cite{Antonelli2021}, which is a biomedical image analysis challenge in which algorithms compete in multiple tasks and image modalities. At the time of writing, the first position on the live leaderboard\footnote{\url{https://decathlon-10.grand-challenge.org/evaluation/challenge/leaderboard/}} is held by the Universal Model \cite{liu2023clip}, an algorithm that uses embedding learned from Contrastive Language-Image Pre-training (CLIP). The second and third positions are held by architectures that use a transformer \cite{Vaswani2017} proposed by \cite{Hatamizadeh2021}. The fourth position is occupied by the nnU-Net \cite{Isensee2020}, a segmentation pipeline based on U-Net that automatically configures to any new medical image segmentation task. Despite state-of-the-art performance on several medical image segmentation tasks, automatic segmentation algorithms have not yet reached the desired robustness to allow clinical use \cite{Sakinis2019}. In particular, segmentation accuracy can be impacted by differences in anatomy per subject, acquisition differences, and image artifacts \cite{Zhao2013}. 



Relatively as compared to automatic segmentation, interactive segmentation methods based on deep learning have been proposed for more robust natural image segmentation \cite{Xu2016, Agustsson2019}. In \cite{Xu2016}, user foreground and background clicks were converted into euclidean distance maps and subsequently added as additional input channels to a CNN. Inspired by the aforementioned studies and other incremental works, interactive methods for medical image segmentation based on deep learning have been recently proposed \cite{LUO2021, Wang2018, Sakinis2019, Wang2019}. In \cite{Wang2018}, a bounding-box and scribble-based network segmentation pipeline was proposed, whereby an initial segmentation is obtained within a user-provided bounding box, followed by image-specific fine-tuning using user-provided scribbles. In contrast, \cite{LUO2021, Sakinis2019} proposed a click-based method, motivated in part by the work of \cite{Xu2016}. In their work, Gaussian-smoothed foreground and background clicks were added as input channels to an encoder-decoder CNN. Experiments on multiple-organ segmentation on CT showed that their proposed method delivers 2D segmentation in a fast and reliable manner, generalises well to unseen structures, and produces accurate results with few clicks. An alternate method that first performs an automatic CNN segmentation step, followed by an optional refinement through user clicks or scribbles, was proposed by \cite{Wang2019}. Their method, named DeepIGeoS, achieved substantially improved performance compared to automatic CNN on 2D placenta and 3D brain tumor segmentation, and higher accuracy with fewer interactions compared to traditional interactive methods.

Automatic and semi-automatic segmentation methods are available as part of open-source software packages for medical image and biomedical image analysis. For instance, ITK-SNAP \cite{py06nimg} offers semi-automatic active contour segmentation \cite{Kass1988}; 3D Slicer \cite{Fedorov2012} and MITK \cite{Nolden2013} offer automatic, boundary-points-based \cite{Maninis2018,roth2021going}, and DeepGrow \cite{Sakinis2019} offers positive and negative interaction click based segmentation through the NVIDIA Clara AI-Assisted Annotation Extension \cite{NVIDIAIAA}, as well as other semi-automatic segmentation methods such as region growing \cite{Adams1994} and level sets \cite{Osher1988}. 

On the other hand, biomedical image analysis tools such as ilastik \cite{ilastik2019} facilitate the use of machine learning algorithms on tasks such as segmentation, object detection, counting, and tracking. However, ilastik does not include the option of training deep convolutional networks, which puts this system at disadvantage when compared to other tools that do use deep learning algorithms. Another popular open-source platform for biomedical image analysis is BioMedisa \cite{Losel2016, Losel2020}, an online platform developed for the semi-automatic segmentation of large volumetric images. It also offers different GPU-based algorithms and weighted random walks for smart interpolation of pre-segmented slices. Biomedisa has significant advantages over CPU-based semi-automatic segmentation tools for biomedical image analysis. It was specifically developed to work on a cluster or parallel computer architectures. This means, that using Biomedisa with a single and small GPU is not always easy.


In addition to open-source software packages, automatic and semi-automatic methods for annotating images are available in commercial solutions such as SuperAnnotate\footnote{\url{https://superannotate.com/}}, V7 Labs\footnote{\url{https://www.v7labs.com/}} and Segments\footnote{\url{https://segments.ai}}. However, the details of the algorithms behind these platforms are not open-source, which makes them less attractive to the scientific community.

\begin{figure*}
\centering
\includegraphics[width=0.95\textwidth]{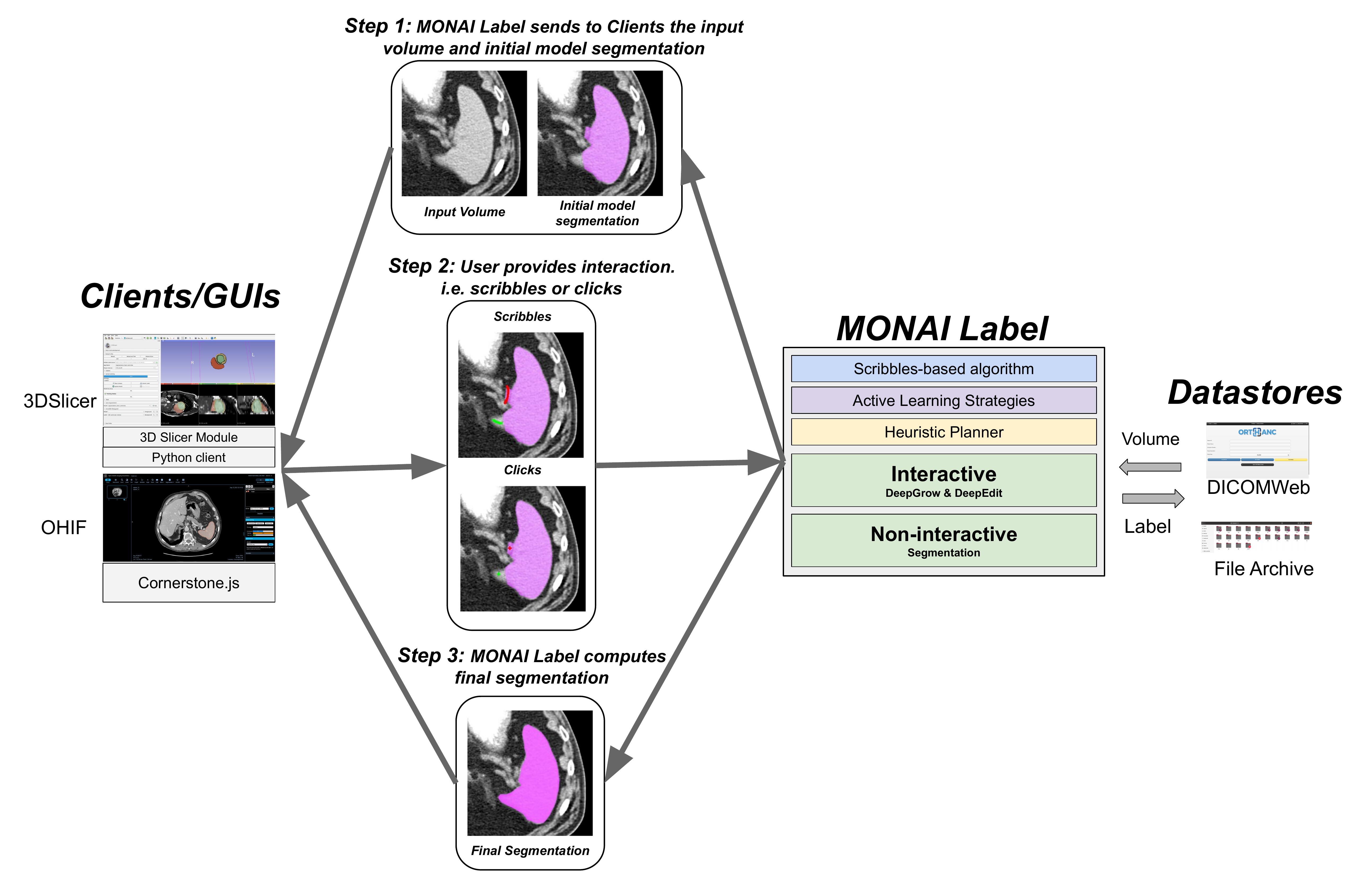} 
\caption{\textbf{Overview of the MONAI Label platform:} MONAI Label consists of three main high-level modules, namely (i) client, (ii) server, and (iii) Database/Datastores. On the client side, MONAI Label supports different graphical user interfaces (GUIs) (3D Slicer/OHIF) for data viewing and annotation. On the server side, AI-assisted interactive and non-interactive annotation methods}. The server also includes active learning strategies and a heuristic planner for improving the efficiency of the underlying annotation methods. Studies provide a database of the dataset that is to be annotated, as well as to store the annotations.
\label{schema_monailabel}
\end{figure*}

MONAI Label is a free and intelligent open-source image labeling and learning framework that enables users to create annotated datasets and build AI-based annotation models for clinical evaluation in a labeling app form. As an SDK, MONAI Label enables researchers to build their own labeling apps, and train and perform inference using deep neural networks. The labeling apps are exposed as a service through the MONAI Label Server. MONAI Label currently offers two types of segmentation approaches: an interactive and a non-interactive approach. The interactive approach is composed of two deep-learning-based models (DeepEdit \cite{Diaz-Pinto-DeepEdit} and DeepGrow \cite{Sakinis2019}) and the scribbles-based method that relies on energy-based optimization \cite{Boykov2006, Wang2018, criminisi2008geos}, and the non-interactive approach is the typical model based on any type of convolutional or transformers-based architecture. MONAI Label also includes a heuristic planner, an algorithm that enables better utilization of available GPU hardware by proposing the best configurations possible for training/inferring from an AI-based model. The heuristic planner can be used in conjunction with any available AI-based annotation techniques. An overview of the different segmentation approaches offered by MONAI Label is shown in Fig. \ref{schema_monailabel}.

Similar to popular annotation platforms for 2D images (Visual Object Tagging Tool (VoTT)\footnote{\url{https://github.com/microsoft/VoTT}} and Computer Vision Annotation Tool (CVAT)\footnote{\url{https://github.com/openvinotoolkit/cvat}}), MONAI Label offers active learning strategies to facilitate the training of deep learning algorithm for 3D medical image segmentation. Specifically, it computes aleatoric and epistemic uncertainty values to rank unlabeled images, allowing the clinician/user to segment the harder samples first. It also communicates through the family of DICOM REST APIs for retrieval of files (WADO-RS), storage (STOW-RS), and querying collections (QIDO-RS) enabling users to integrate MONAI Label in their Picture Archiving and Communication System (PACS), XNAT \cite{Marcus2007}, Image Data Commons (IDC) \cite{Fedorov2021} or any other DICOM system. This feature allows web developers to unlock the power of healthcare images using industry-standard toolsets. Additionally, MONAI Label supports any graphical user interfaces (GUI) talking to its Rest API. As an example, there are currently two integrations with two popular GUIs: 3D Slicer and OHIF. See Fig. \ref{schema_monailabel}.


Documentation and MONAI Label application examples can be found at \url{https://pypi.org/project/monailabel/}. It can be installed from the Python Package Index (PyPI) by running \textbf{\textit{pip install monailabel}}, which also installs the pre-built OHIF viewer \cite{urban2017ohif}.


\section{Annotation Approaches}
\label{annotation_approaches}

MONAI Label currently offers two annotation approaches: an interactive approach consisting of two deep-learning-based models (DeepGrow, DeepEdit) and one that relies on energy-based optimization (Scribbles-based method). The other approach is the non-interactive (Automatic Segmentation) which is the typical deep learning segmentation model. Both DeepGrow and DeepEdit guide the segmentation with clicks provided by the user while the scribbles-based method utilizes scribbles provided by the user to guide the segmentation. On the other hand, the non-interactive approach allows the researcher to create a segmentation pipeline using a segmentation network implemented in MONAI or the underlying PyTorch framework (i.e. UNet \cite{Ronneberger2015}, Highresnet, ResNet \cite{He2016}, DynUnet \cite{MONAI2020}, etc ) to automatically segment images.

\subsection{Interactive Approaches}
\subsubsection{DeepGrow Model}

DeepGrow is an interactive segmentation model where the user guides the segmentation with positive and negative clicks. The positive clicks, inside the region of interest, expand the segmentation to include that location, while the negative clicks are used for contracting the segmentation to exclude the clicked region from the region of interest \cite{Sakinis2019}.

The training process of a DeepGrow model differs from traditional deep learning segmentation due to a simulation process of positive and negative guidance (clicks) involved in the training process. The positive and negative guidance maps are generated based on the false negatives and false positives which are dependent on the predictions. Both DeepGrow 2D and 3D allow the user to annotate only one label at a time. DeepGrow 2D allows the user to annotate the image one slice at a time, whereas DeepGrow 3D can annotate whole volumes.

\begin{figure*}
\centering
\includegraphics[width=0.9\textwidth]{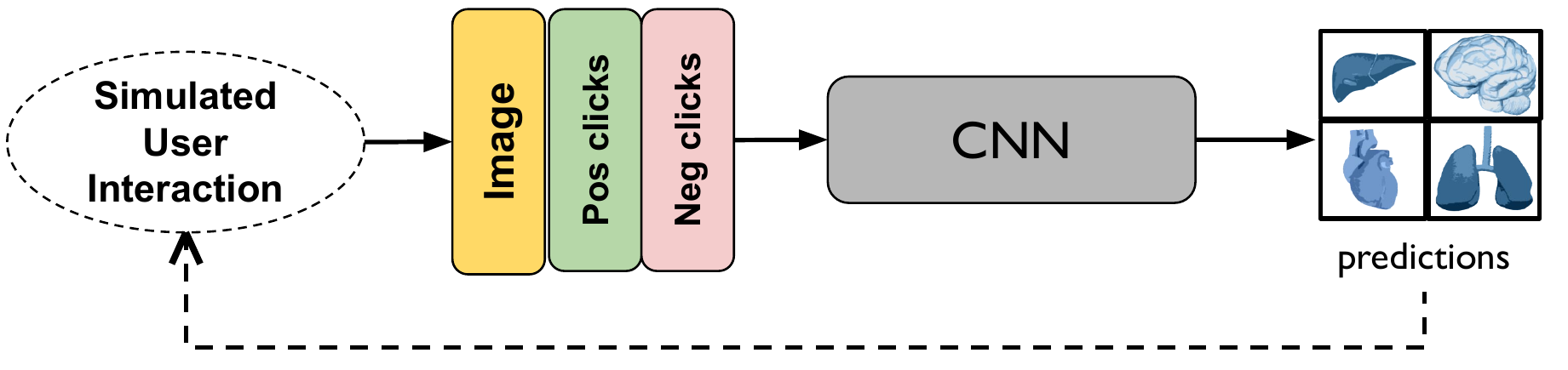} 
\caption{\textbf{Schema of the DeepGrow approach:} The input tensor consists of the image and two tensors representing positive (Pos) and negative (Neg) clicks provided by the user.}
\label{schema_deepgrow}
\end{figure*}

\subsubsection{DeepEdit Model}

DeepEdit extends DeepGrow's click-based segmentation to allow for click-free segmentation inference and click-based segmentation editing \cite{Diaz-Pinto-DeepEdit}. More specifically, it allows the user to perform inference, as a standard segmentation method (i.e. UNet), and also to interactively segment part of an image using clicks as DeepGrow does \cite{Sakinis2019}. DeepEdit facilitates the user experience and development of new active learning \cite{gal2017deep} techniques as no user interaction (i.e. clicks) is needed to obtain an initially predicted mask, thus allowing the integration of active learning strategies to prioritize the labeling process.

\begin{figure*}
\centering
\includegraphics[width=0.9\textwidth]{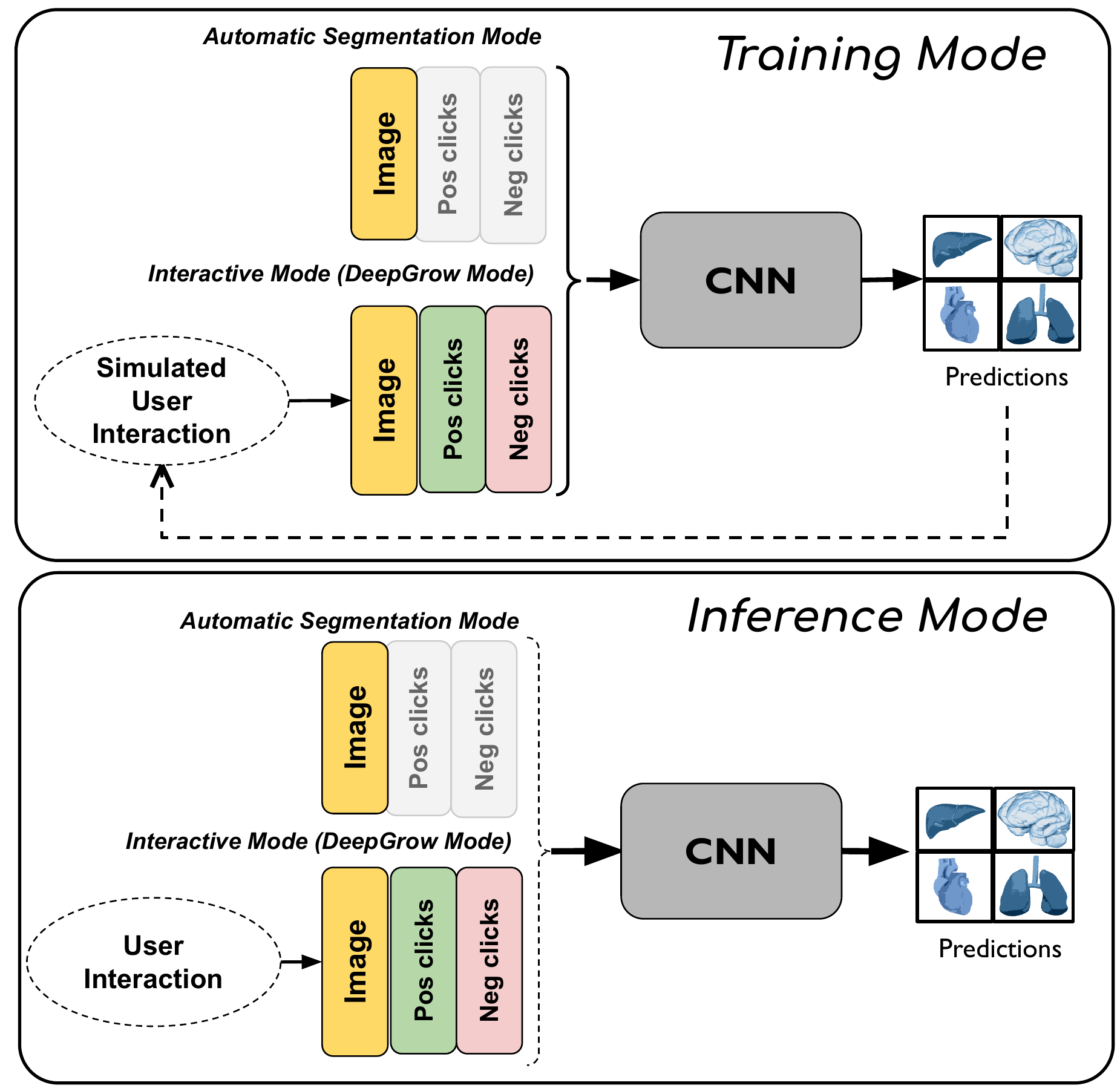} 
\caption{\textbf{General schema of the DeepEdit approach:} DeepEdit training process consists of a combination of two modes: the automatic segmentation mode and DeepGrow mode. For inference, the input tensor could be either the image with two zero-tensor (automatic segmentation mode) or an image with two tensors representing positive and negative clicks provided by the user (interactive mode or DeepGrow mode).}
\label{schema_deepedit}
\end{figure*}

The training process of the DeepEdit approach involves a combination of simulated clicks and standard non-interactive training. As shown in Fig. \ref{schema_deepedit} (Training Mode), the input of the network is a concatenation of three tensors: image, positive clicks representing the foreground, and negative clicks representing the background. This model has two training stages: For half of the iterations, tensors representing the foreground and background points are zeros and for the other half, positive and negative clicks are simulated following the process presented by \cite{Sakinis2019}. 

For automatic inference, the tensors representing positive and negative clicks are replaced by zeros. However, for the interactive segmentation mode, positive and negative points/clicks provided by the user are placed in the channels accordingly. See Fig. \ref{schema_deepedit} (Inference Mode)

\subsubsection{Scribbles-based Segmentation Model}
Scribbles are free-hand drawings, such as drawing with a pen on paper, which have been widely employed to propose a range of interactive segmentation methods \cite{Wang2018, boykov2004experimental, rother2004grabcut}. Scribbles provide natural interaction, which most annotators are already familiar with. These interactions introduce flexibility in annotators' workload, i.e. it can be as involved as required; providing both minimal interactions for simpler delineation tasks and detailed interactions for more difficult segmentations.

\begin{figure*}
\centering
\includegraphics[width=1.0\textwidth]{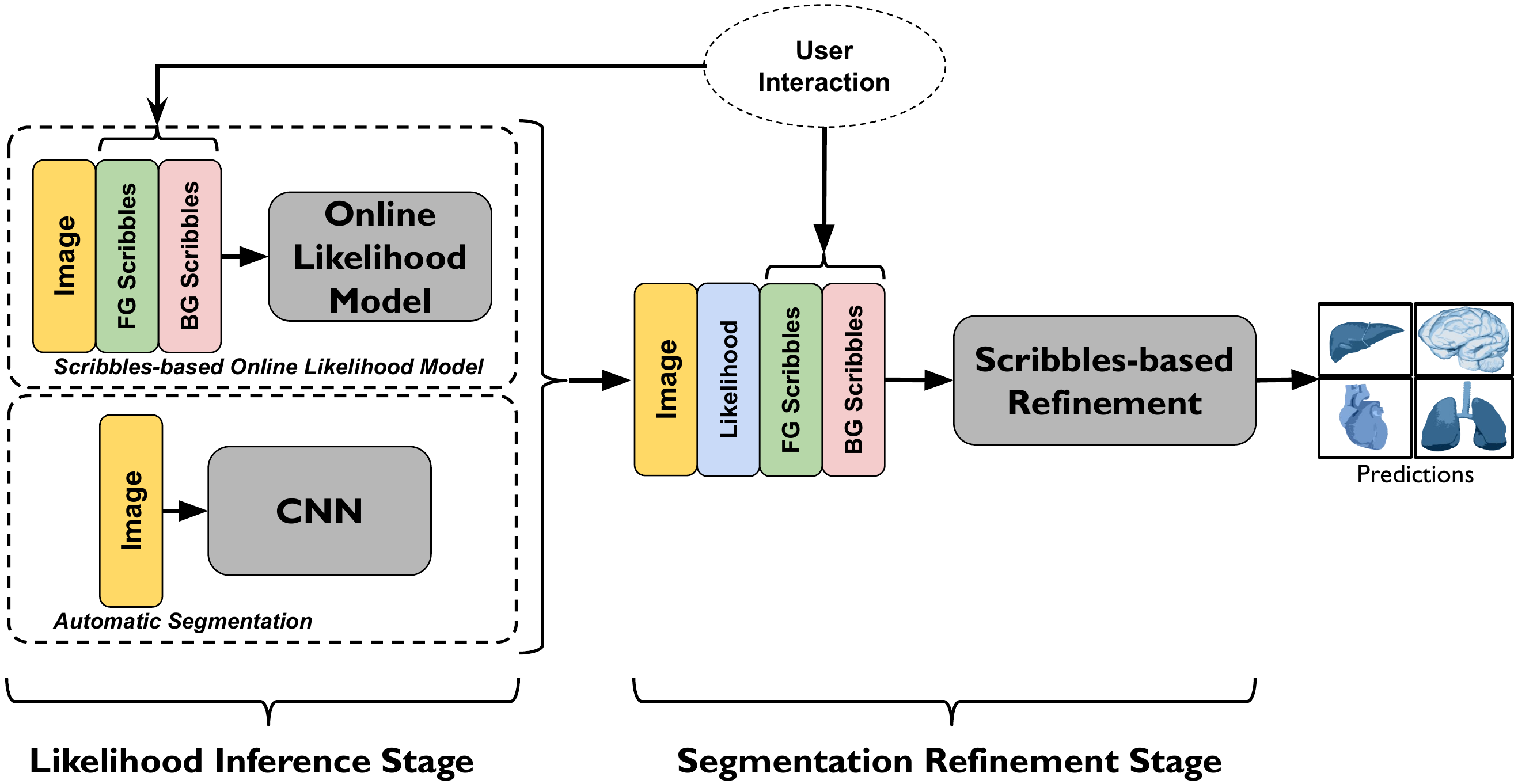} 
\caption{\textbf{Scribbles-based interactive segmentation in MONAI Label:} Scribbles-based methods consist of two stages, namely (i) likelihood inference stage and, (ii) segmentation refinement stage. The likelihood can come from either an online model built using the image volume and user scribbles, or using a pre-trained deep learning model on image volume alone. The interactions are provided as foreground (FG) or background (BG) scribbles. Image volume, likelihood, and scribbles are then used in a refinement stage to refine the initial segmentation using an energy optimization approach, e.g. using GraphCut \cite{boykov2004experimental}.}
\label{fig:scribbles-overview}
\end{figure*}

MONAI Label provides APIs for implementing scribbles-based interactive segmentation workflows. An overview of such workflows is presented in Fig. \ref{fig:scribbles-overview}.

In the figure, it can be seen that the scribbles-based interactive segmentation can be used in two different modes which are:
\begin{enumerate}
    \item Scribbles-based online likelihood segmentation: uses scribbles to generate segmentation labels
    \item Scribbles-based CNN segmentation refinement: refines segmentations from a deep learning model using user-scribbles
\end{enumerate}
MONAI Label provides sample applications for both 1. and 2.

More information regarding the Scribbles-base segmentation model can be found in the Supplementary Material

\subsection{Non-interactive Approach}
\subsubsection{Automated Segmentation Model}

The MONAI Label supports automated segmentation as an annotation approach which is essentially a non-interactive algorithm based on a standard convolutional neural network (CNN) (i.e. UNet). Researchers can use any of the available networks created in MONAI Core for their purposes. In terms of classes and interfaces, this application follows the same structure as the DeepGrow and DeepEdit approaches. The most important difference between this annotation app and the interactive ones is the extra information added to the input images. In this app, extra channels representing foreground and background clicks are not added to the input image. Note that the automated segmentation can be then further edited manually using standard segmentation tools provided in the clients.

\section{Active Learning}

\begin{figure*}
\centering
\includegraphics[width=0.8\textwidth]{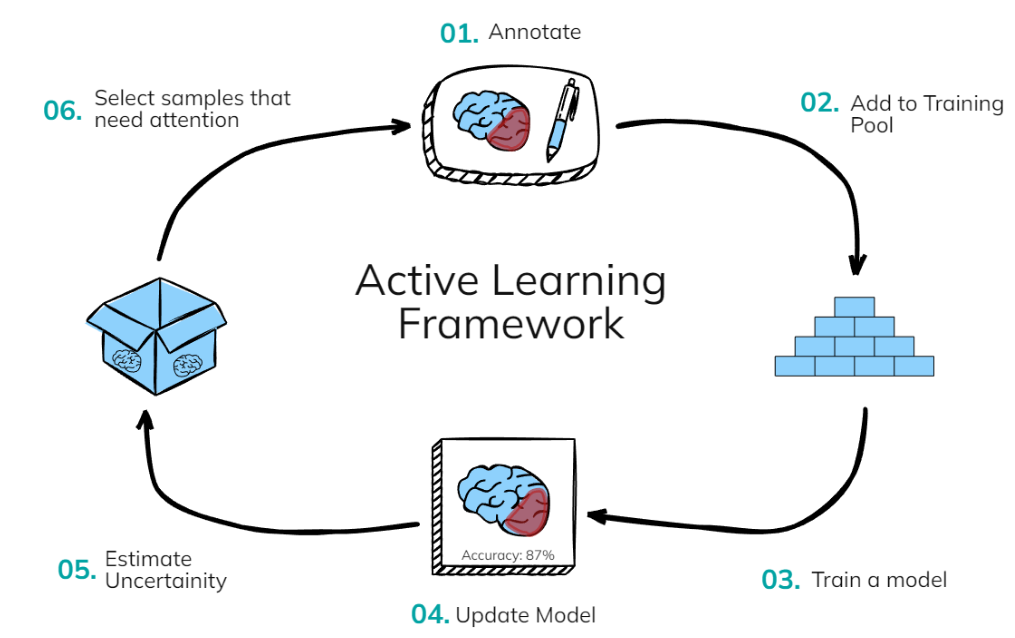} 
\caption{\textbf{Active learning cycle:} An expert labeller annotates a volume and adds it to the training pool. A machine learning model is trained with the new annotation obtaining a model. This model allows for uncertainty estimation which is utilized to rank the unlabeled volumes in order to select a set of the most uncertain volumes. The selected uncertain data are labeled by the expert which are labeled and added to the training pool and the cycle can be repeated as many times till a model of desired performance is achieved.}
\label{schema_active_learning}
\end{figure*}

Active learning (AL) is a powerful data-selection engine as it targets the \textit{most relevant} data for the training of the model, with the annotation help from an expert the data can be utilized towards the training of the model (See Fig. \ref{schema_active_learning}). AL was initially heavily advocated for by \cite{settles2012active} and it was applied with many classical machine learning algorithms. In recent times deep learning has given it additional attention as data has become the key for powerful models and AL provides the framework to get the \textit{relevant} data for models. AL has multiple components in the framework: the machine learning model, the unlabeled pool of data, the acquisition function and the uncertainty estimation of the data point. All the components have their own hyper-parameters that drive the performance of the AL framework.

\textit{Uncertainty Estimation}: There are many established ways to derive uncertainty from a machine learning model. Query-by-committee \cite{nath2020diminishing} which is an ensemble system of models and their disagreements serve as the uncertainty measure. Utilizing dropout \cite{gal2017deep, nath2020diminishing} during inference with bootstrapping is a way to estimate uncertainty using a single model. There are also bayesian models that directly offer uncertainty estimates as the parameters of the model itself are defined using mean and standard deviations. Currently, MONAI Label supports dropout-based uncertainty strategies, the motivation being that a single model is relatively cheaper to compute as compared to an ensemble of models. It should be noted though that a user of MONAI Label can create an application to utilize ensemble or bayesian models if they prefer it.

\textit{Acquisition Functions: } Estimation of uncertainty is for a particular data point however selection of data is dependent upon acquisition function. For e.g the simplest acquisition function is to select a subset of the most uncertain data points. However, there are many other complex functions such as maximum cover \cite{yang2017suggestive}, knapsack \cite{kuo2018cost} or combinations of uncertainty and data representativeness via mutual information \cite{nath2020diminishing}. Many comprehensive studies list a more detailed and elaborate comparison of acquisition functions \cite{zou2023review}. MONAI Label provides the most commonly used acquisition functions such as the selection of the most uncertain volumes, however, a user can define their own acquisition function in a new application.

\textit{Network Architecture:} Dropout-based techniques are model agnostic and therefore as long as dropout is being used uncertainty can be generated.

\textit{Hyperparameters of AL:} Some common hyper-parameters that drive AL are the number of simulations when using dropout for estimation of uncertainty, the number of data points to be annotated or annotation budget. Other parameters that are particular to acquisition functions themselves etc.

It is worth highlighting that as DeepEdit is a combination of two models: a non-interactive and an interactive one, it allows the usage of uncertainty-based AL techniques. The non-interactive part of DeepEdit can be used to run inference on unlabelled images and sort them using uncertainty computation algorithms. This is different from the DeepGrow model which needs clicks to compute a predicted mask and then compute uncertainty on the unlabelled images.

\section{Graphical User Interfaces}

\subsection{Locally-installed: 3D Slicer}

3D Slicer is an open-source multi-platform software package widely used for biomedical and medical imaging research \cite{Fedorov2012}. The module implemented to work with MONAI Label handles calls/events created by the user interaction. The current version supports click interaction and allows the user to upload images and labels. Additional interactions such as closed curves, ROI or any other are supported by the MONAI Label server. Researchers can modify this module to make it more dynamic with respect to the type of interaction, and available buttons to consume the MONAI Label REST API or to customise their MONAI Label segmentation applications.


\subsection{Web-based: Open Health Imaging Foundation (OHIF)}

The OHIF Viewer is an open-source and web-based viewer \cite{urban2017ohif}. It is based on Cornerstone.js and works out-of-the-box with Image Archives that support DICOMWeb. MONAI Label has OHIF embedded and it works with the DICOMWeb server support. It also allows the user to create their own label mask and interact with the MONAI Label server. Also important to note that OHIF has been integrated into XNAT and other large initiatives/centers, meaning that it would be easier to integrate into complex workflows.


\section{Developing and Deploying MONAI Label Apps}

The MONAI Label framework empowers researchers with the ability to develop novel annotation methods and make them readily available to other researchers and clinical staff for continuous evaluation of user experience and annotation performance. Developing a MONAI Label App requires the implementation of a simple Python API that defines what models are used during annotation, how they learn from user interactions, and how active learning is employed to shorten annotation time as work progresses.

MONAI Label Apps are exposed as a service using the MONAI Label Server. The MONAI Label server exposes the functionality in the MONAI Label App via a RESTful API.


MONAI Label Server enables data management for all MONAI Label Apps, allowing app developers to track various metadata related to training, AI-based annotation, and active learning, and provides caching for images and labels. Data management in MONAI Label Server supports both locally stored datasets and data residing in DICOM servers (accessible via DICOMWeb).







\subsection{Application Call-flow}

MONAI Label App developers may choose to implement part or all of the following functionality in their apps.

\begin{itemize}
    
    \item \textbf{Active Learning:} MONAI Label offers all the necessary callbacks needed to implement custom active learning techniques. Natively, MONAI Label supports simple active learning \cite{top2011active, nath2020diminishing} techniques such as in-sequence and random sampling, and more highly performant active learning techniques such as Test Time Augmentation \cite{Wang2019TTA} (aleatoric uncertainty) and Montecarlo computation using Dropout (epistemic uncertainty) \cite{gal2017deep}. 
    

    \item \textbf{Inference:} This interface allows the developer to use or define the inference task in the MONAI Label App which will define the behavior of the user interaction (i.e. clicks, scribbles, ROI, etc). The developer will need to define the inference models and inferrers (simple inferrer, sliding window inferrer, etc), and pre-and post-transform.

    \item \textbf{Training:} This interface allows the developer to define the training task for their own MONAI Label App which may include one or more models with the appropriate pre- and post-processing in place. Here, the developer also decides the best data splitting strategy and validation metrics.

    \item \textbf{Custom transforms:} In this file developers can create their custom transforms to apply to the dataset before, after, or during training the deep learning algorithm.

    \item \textbf{Main Module:} In this file is where developers define the core structure of the MONAI Label App and instantiate the modules they defined above.

    \item \textbf{Requirements File:} Here, researchers have the option of specifying the external libraries required by their app (e.g. TensorFlow, Catalyst, Kornia \cite{eriba2019kornia}, etc)

\end{itemize}



MONAI Label enables researchers to build labeling applications, ready to deploy via its API. To develop a new MONAI labeling app, developers must inherit the \texttt{MONAILabelApp} interface and implement the methods in the interface that are relevant to their application. Typically a labeling application will consist of:

\begin{itemize}
    \item Inferencing tasks to allow end-users to invoke selected pre-trained or actively trained models.
    \item Training tasks used to train a set of models in the background and perhaps without the users' participation to actively improve annotation as work progresses, 
    \item Image selection strategies that choose the unannotated image that is least represented in the already labeled images.
\end{itemize}

\begin{figure*}
\centering
\includegraphics[width=0.8\textwidth]{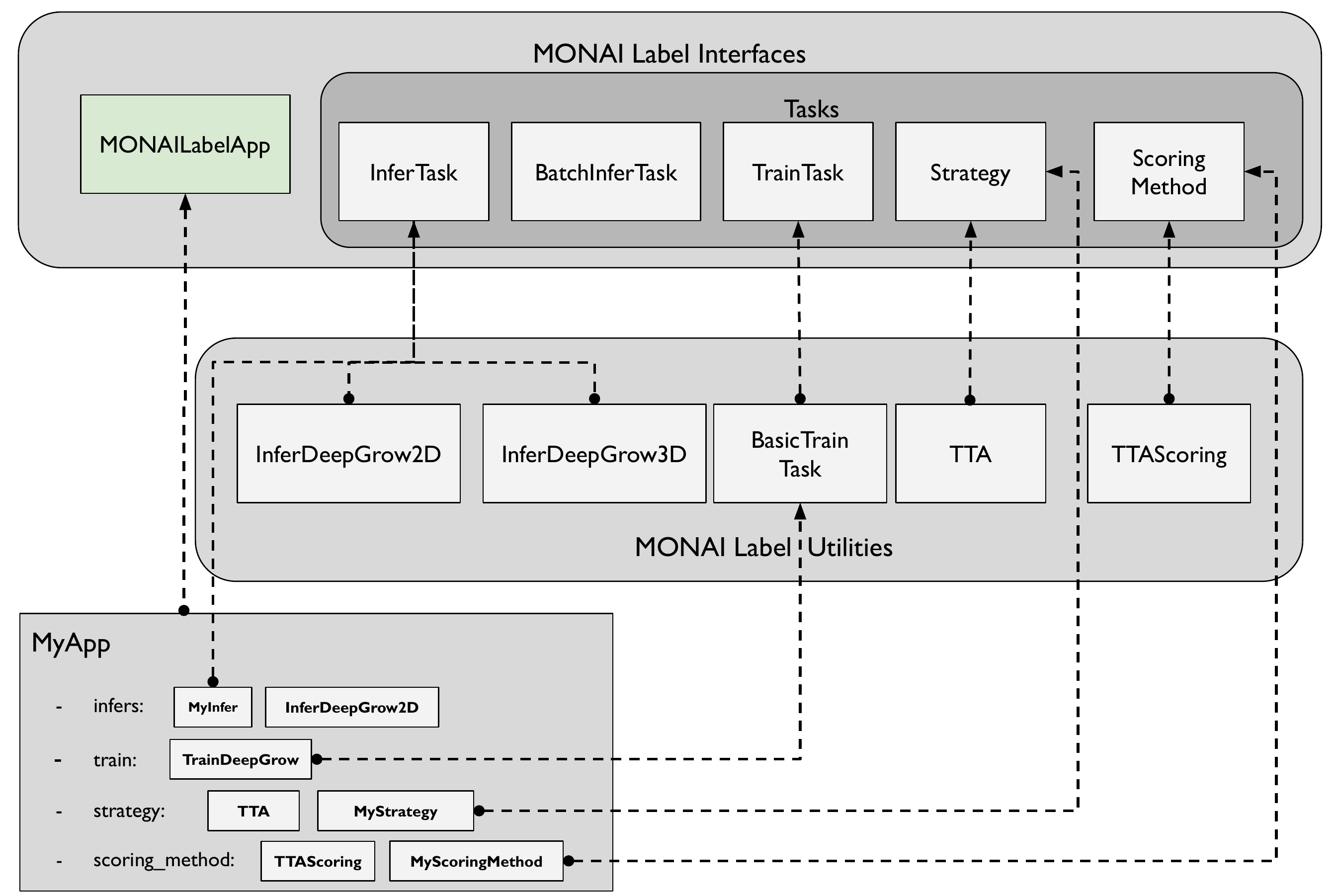} 
\caption{\textbf{Modules Overview:} MONAI Label provides interfaces that can be implemented by the label app developer for custom functionality as well as utilities that are readily usable in the labeling app.}
\label{monailabel_base_interfaces}
\end{figure*}

Fig. \ref{monailabel_base_interfaces} shows the base interfaces that a developer may use to implement their app and the various tasks their app may perform. For example, in the figure the app \texttt{MyApp} employs: two inferencing tasks, namely \texttt{MyInfer}, which is a custom implementation of \texttt{InferTask}, and \texttt{InferDeepGrow2D}, which is a ready-to-use utility included with MONAI Label. There is also one training task, \texttt{TrainDeepGrow} which is an extension of the \texttt{BasicTrainTask} utility, and two ``next image selection'' strategies, \texttt{TTA} and \texttt{Epistemic} included with MONAL Label. These two strategies allow the user to select the next image based on the aleatoric or epistemic uncertainty values. Finally, there is also \texttt{MyStrategy} which implements the interface \texttt{Strategy} which the end user may select as a custom alternative for next image selection.

MONAI Label currently provides template apps and segmentation approaches that developers may start using out of the box or modify to achieve the desired behavior. Template applications currently include annotations based on DeepGrow \cite{Sakinis2019}, DeepEdit, and automated segmentation. All template apps come with support for Scribbles-based segmentation.

\section{Experiments and Results}

\subsection{DeepGrow Performance}

\textit{Experiment:} The spleen segmentation dataset from the medical segmentation challenge (MSD) was utilized for experimentation. The dataset consists of a total of 41 3D CT volumes. Six volumes were randomly selected for validation and the remaining were divided into five for initial training and the others assumed as unlabeled volumes. The experiment is conducted in a 4 stage approach where at every stage the user annotates a certain number of 3D volumes and then adds them to the training pool. At stage 1 the user annotates 11 volumes, at stage 2 the user adds another 5 and at stage 3 and stage 4 the user annotates 10 volumes (more volumes could be annotated because the user could annotate faster due to AI assistance or deepgrow-based annotation). For comparison, we also evaluated the scribbles-based online likelihood segmentation that aims at minimizing human interactions needed to manually annotate a dataset. Baselines of paintbrush and advanced contour-based techniques were also used for estimating how much time is needed by a human annotator who is utilizing traditional annotation tools. To perform a fair comparison the training time for AI models will be observed separately and will not include in the annotation time when compared to traditional approaches. The reason is that training is performed once and is not a part of the annotation process.

\textit{Results:} Observing Table \ref{table_deepgrow_validation}, it can be seen that the time taken by the user to annotate a single 3D volume grows lesser as more training data is added per stage for the AI model to learn from. At 4$^{th}$ stage it can be observed that by utilizing the combined pipeline of Deepgrow 2D \& 3D the user can annotate 3D volumes in approximately 1 - 2.5 minutes which is 10x faster even compared to the advanced traditional technique of contouring manually annotate the 3D volume \cite{Fedorov2012}.

Comparing scribbles-based online likelihood segmentation with the traditional paintbrush and contour-based techniques in Table \ref{table_deepgrow_validation}, we note that the scribbles-based method significantly improves the time to manually annotate the dataset, while minimizing user interactions required. In particular, on average 2 minutes were required to annotate a sample using the scribbles-based method which is $12.5 \times$ and $6.25 \times$ faster than using the paintbrush and contour-based method, respectively.

\begin{table*}
\caption{\textbf{Time spent on annotating the Spleen MSD dataset:} This table presents the times spent on each stage annotating volumes from the Spleen dataset using several manual tools (paintbrush, contor-based and scribbles-based methods) and the interactive DeepGrow model available on MONAI Label.}
\resizebox{\textwidth}{!}{
\begin{tabular}{|c|c|c|c|c|c|}
\hline
        & \begin{tabular}[c]{@{}c@{}}Annotated\\ Volumes\end{tabular}    & \begin{tabular}[c]{@{}c@{}}Paint Brush\\ Method\end{tabular} & \begin{tabular}[c]{@{}c@{}}Contour-Based\\ Method\end{tabular} & \begin{tabular}[c]{@{}c@{}}Scribbles-Based\\ Method\end{tabular} & \begin{tabular}[c]{@{}c@{}}\textbf{DeepGrow} \\ \textbf{Method}\end{tabular} \\ \hline
Stage 1 & \textbf{11}                      & 275 mins                                            & 137.5 mins                                              & 22 mins                                                 & \textbf{25 mins}          \\ \hline
Stage 2 & 11 + \textbf{(5)} = 16           & 400 mins                                            & 200 mins                                                & 32 mins                                                 & \textbf{6 - 7.5 mins}      \\ \hline
Stage 3 & 11 + 5 + \textbf{(10)} = 26      & 650 mins                                            & 325 mins                                                & 52 mins                                                 & \textbf{3.5 - 5 mins}      \\ \hline
Stage 4 & 11 + 5 + 10 + \textbf{(10)} = 36 & 900 mins                                            & 450 mins                                                & 72 mins                                                 & \textbf{1 - 2.5 mins}      \\ \hline
\end{tabular}
}
\label{table_deepgrow_validation}
\end{table*}

\begin{table*}[h!]
\caption{\textbf{Obtained results from the interactive DeepGrow model:} Total annotation time per stage, training time on each stage, and the validation Dice Scores on the Spleen MSD dataset. The validation set is composed of nine 3D volumes (20\%) that were randomly selected from the Spleen MSD dataset.}
\resizebox{\textwidth}{!}{
\begin{tabular}{|c|c|c|c|c|}
\hline
        & \begin{tabular}[c]{@{}c@{}}Total Annotation Time\\ Using DeepGrow\end{tabular} & \begin{tabular}[c]{@{}c@{}}Training Time\\ DeepGrow 2D \& 3D\end{tabular} & \begin{tabular}[c]{@{}c@{}}Validation Dice\\ DeepGrow 2D\end{tabular} & \begin{tabular}[c]{@{}c@{}}Validation Dice\\ DeepGrow 3D\end{tabular} \\ \hline
Stage 1 & 275 mins             & 90 mins                                                                   & 0.891                                                          & 0.730                                                          \\ \hline
Stage 2 & 30 mins              & 135 mins                                                                  & 0.924                                                          & 0.873                                                          \\ \hline
Stage 3 & 45 mins              & 250 mins                                                                  & 0.948                                                          & 0.945                                                          \\ \hline
Stage 4 & 15 mins              & 360 mins                                                                  & 0.967                                                          & 0.959                                                          \\ \hline
\end{tabular}
}
\label{deepgrow_validation_time_score}
\end{table*}

\subsection{DeepEdit Performance}

In order to demonstrate how the DeepEdit annotation approach can be used to facilitate medical image segmentation, we used cardiac magnetic resonance images (CMR) from the Cardiac task available in the Medical Segmentation Decathlon (MSD) \cite{Antonelli2021} to segment the left atrium. This dataset is composed of 20 CMR images. A split of 80\% for training and 20\% for validation was used in this example. This means, 16 images were randomly selected for training and 4 images for validation. A learning rate of 1e-4, batch size equal to 1, Adam optimizer, 50 epochs by default, and random affine transformation were used for data augmentation.

\begin{figure*}
\centering
\includegraphics[width=\linewidth]{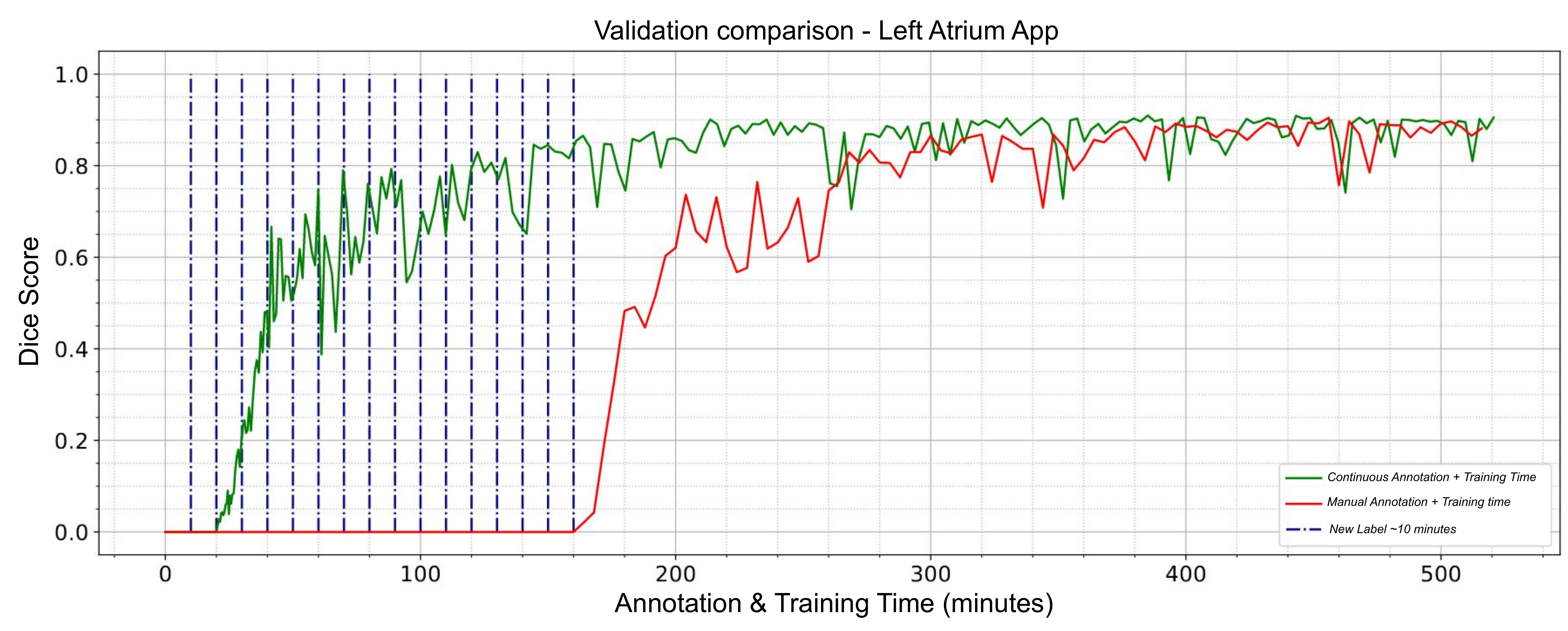} 
\caption{\textbf{Validation of the DeepEdit approach:} Obtained results for left atrium segmentation on the MSD dataset using the DeepEdit model (interactive approach). Vertical blue lines represent the time spent by a clinician to annotate and submit a label to the MONAI Label platform (approx. 10 minutes per volume). The green line represents the Dice score obtained from the interactive approach. The red line represents the Dice score obtained from the manual annotation approach. X-axis (Time in minutes) represents both annotation (continuous or manual) and the training time. By using DeepEdit via MONAI Label, users can train and annotate at the same time.}
\label{left_atrium_app_val}
\end{figure*}

For this experiment, we measured the time an expert annotator took to manually annotate the left atrium (10 minutes) using the manual/basic available tools in 3D Slicer (i.e. Grow from Seeds, Brush, etc). This means, that a clinician should spend around 160 minutes to fully segment the train split (16 CMR images) before they can start training a deep learning model (See red line in Fig. \ref{left_atrium_app_val}). However, if the clinician uses MONAI Label with the DeepEdit approach, they could start the training process after segmenting the first one or two CMR images. This allows the clinician to use the obtained model to continue the annotation of the other images. 

The model trained on one or two images might not perform well in the beginning, but it helps the clinician to quickly create a label that they can modify using the interactive part of the DeepEdit (clicks), which significantly reduces the time they spend on the other images. See Fig. \ref{left_atrium_app_val} (Green line)

This way of annotating and training interactive deep learning models allows clinicians to reduce the time and effort spent on this process. As it can be seen from Fig. \ref{left_atrium_app_val} and Table \ref{deepgrow_validation_time_score}, MONAI Label could help clinicians to significantly reduce annotation time. This can be seen in the performance improvement in lesser time when comparing the green line [Continuous Annotation + Training Time] against the red line [Manual Annotation of all data + Training time].

\section{Latest MONAI Label Use Cases}

\subsection{Telestroke Service}

\begin{figure*}
\centering
\includegraphics[width=0.9\textwidth]{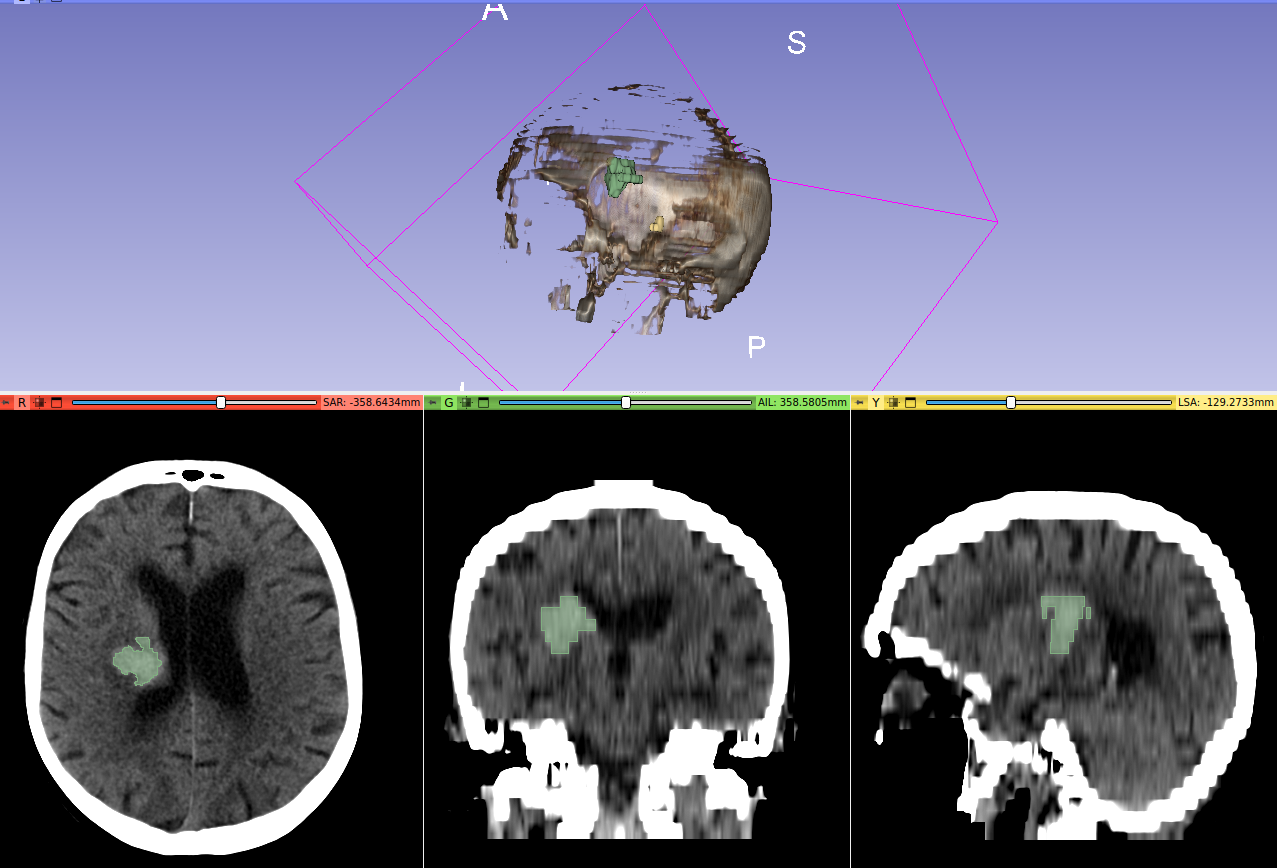} 
\caption{Predicted segmentation mask of brain hemorrhage computed by a MONAI Label model.}
\label{stroke_sample}
\end{figure*}

MONAI Label has been utilised for brain hemorrhage segmentation in the AWS-eHealth cloud (Prince of Wales Hospital, Australia). A customised model has been developed for automated quantitative labelling of CT data. Initial model training has been completed with the MONAI Label+3DSlicer setup and using the available active learning techniques. See label prediction by MONAI Label in Fig. \ref{stroke_sample}.

Once fully trained, the model will be used to automatically identify and measure lesion volumes on images as they are being acquired during acute stroke assessment and a follow-up assessment. This will allow Telestroke clinicians to view images and make assessments as they do now with the added functionality of automated lesion identification and quantification. Clinicians will continue improving the model by approving or rejecting the label provided during the Telestroke consultation using the 3DSlicer plugin in MONAI Label. These re-labelled images will then be used to continuously train the models. This process will be incorporated into the weekly clinical review meeting, where all cases including imaging are discussed by the entire service (consultants and stroke coordinators at each site) using the Microsoft Teams platform.

\subsection{Neurosurgical Atlas}

MONAI Label has been used to create digital twins of diseased brains for surgical planning. The Neurosurgical Atlas team (\url{https://www.neurosurgicalatlas.com/}) has used MONAI Label to segment brain tumors and the surrounding anatomical structures deformed by the tumors based on multimodality MR images (T1, FLAIR, T2, and T1 contrast). A total of 220 lesions were segmented, and divided into meningiomas, low-grade gliomas, high-grade gliomas, and brain metastases, which comprise the most common brain tumors in the general population.

Using the active learning strategies and the non-interactive model available in MONAI Label, neurosurgeons could reduce the time from 50\% to 80\% annotating diseased brains for surgical planning purposes.

\begin{figure*}
\centering
\includegraphics[width=0.8\textwidth]{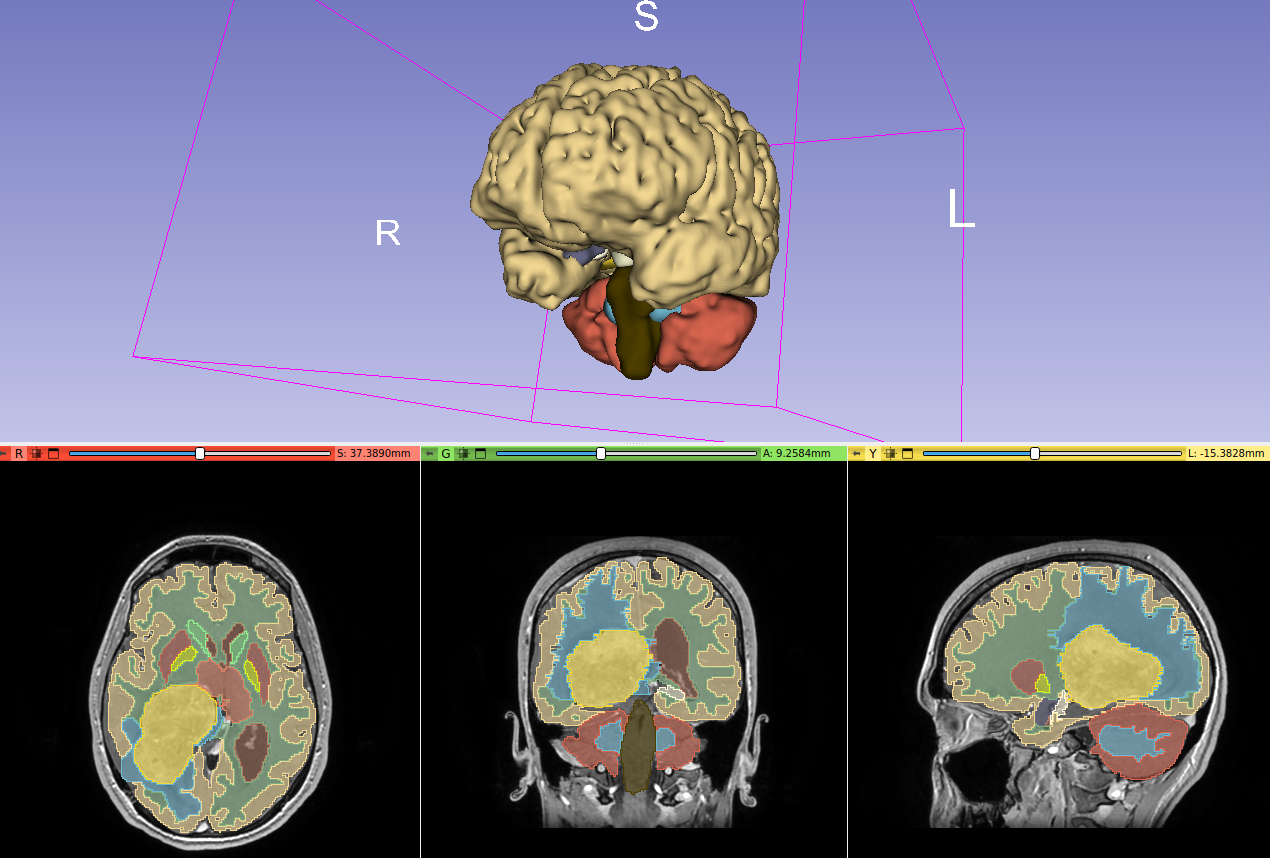} 
\caption{Predicted segmentation mask of brain tumor and deformed brain regions using MONAI Label.}
\label{neuro_atlas_sample}
\end{figure*}

The segmentations obtained from models trained in MONAI Label were later registered to The Atlas for surgery planning using technology designed by the Neurosurgical Atlas. In Fig. \ref{neuro_atlas_sample} it is possible to see the obtained segmentations from MONAI Label). The president of Neurosurgical Atlas, Dr Aaron A. Cohen-Gadol, presented this work during the GPU Technology Conference (GTC) in September 2022 in a session called New Frontiers in Brain Surgery (\url{https://www.nvidia.com/en-us/on-demand/session/gtcfall22-a41130/}).

\section{Conclusion and Future Work}
\label{conclusion_section}

We have introduced MONAI Label\footnote{\url{https://github.com/Project-MONAI/MONAILabel}}, a free and open-source image labeling and learning platform that enables users to create annotated datasets and build AI annotation models for clinical evaluation. MONAI Label reduces the time and effort of annotating new datasets and enables the adaptation of AI to the task at hand by continuously learning from user interactions via two different user interfaces: 3D Slicer and OHIF.

MONAI Label offers two annotation approaches: an interactive approach with methods such as DeepGrow, DeepEdit, and Scribbles-based, and one non-interactive (Automatic Segmentation) which is the standard deep learning segmentation method.

MONAI Label platform is compatible with other libraries beyond MONAI Core~\cite{MONAI2020}. It was developed to help researchers and clinicians to facilitate 3D medical image annotations and allow them to easily implement new machine learning or deep learning algorithms. For that reason, we designed it in a way that users can seamlessly work with different deep learning frameworks and high-level libraries such as Ignite and PyTorch Lightning.

In order to improve integration, we have created an XNAT datastore management interface and the Pathology App which has example models to do both interactive and automated segmentation over pathology (WSI) images - including nuclei multi-label segmentation for neoplastic cells, inflammatory, connective/Soft tissue cells, dead cells, and epithelial.

Finally, MONAI Label offers a heuristic planner that considers available GPU and intensity and spatial information of the training set to define the data transformations and hyperparameters used during training and inference.

Further work includes more user interactions such as ROIs and/or closed curves for all the interactive models. The authors welcome constructive feedback, feature requests, and contributions from the community.


\section*{Code and Data Availability Statements}

MONAI Label is an open source project and is freely available at \url{https://github.com/Project-MONAI/MONAILabel} under the Apache-2.0 license. It was developed and tested for Linux and Windows operating systems. Any common browser can be used to work on OHIF viewer. MONAI Label can be installed according to the installation instructions available in the GitHub project. By the time this article was submitted for review, the authors were working on version 0.3.0 of MONAI Label. \url{https://github.com/Project-MONAI/MONAILabel/releases/tag/0.3.0}

Images used to demonstrate how MONAI Label reduces the time and effort segmenting medical images are publicly available at \url{http://medicaldecathlon.com/}.

\section*{Conflict of Interest} S.O. and T.V. are co-founders and shareholders of Hypervision Surgical. T.V. holds shares in Mauna Kea Technologies. 


\section*{Acknowledgements}
A.D.-P. is supported by the London AI centre. This work was partially supported by European Union's Horizon 2020 research and innovation programme under grant agreement No 101016131 (icovid project). This work was supported by core and project funding from the Wellcome/EPSRC [WT203148/Z/16/Z; NS/A000049/1; WT101957; NS/A000027/1]. T.V. is supported by a Medtronic / RAEng Research Chair [RCSRF1819\textbackslash7\textbackslash34].

\bibliographystyle{model2-names.bst}\biboptions{authoryear}
\bibliography{references.bib}

\end{document}